\documentstyle[epsf,aps,multicol]{revtex}
\begin{document}
%\psdraft
\draft
\renewcommand\topfraction{0.1}
\title{Statistical reconstruction of three-dimensional porous
media from two-dimensional images}
\author{Anthony P. Roberts$^{\small *}$}
\address{
Faculty of Environmental Sciences, Griffith University, Nathan,
Queensland 4111, Australia}
\date{15 April 1997}
\maketitle
\begin{abstract}
A method of modelling the three-dimensional microstructure of random
isotropic two-phase materials is proposed.
The information required to implement the technique can be obtained
from two-dimensional images of the microstructure.
The reconstructed models share two-point correlation and chord-distribution
functions with the original composite.
The method is designed to produce models for computationally
and theoretically predicting the effective macroscopic properties of
random materials (such as electrical and
thermal conductivity, permeability and elastic moduli).
To test the method we reconstruct the morphology and predict the
conductivity of the well known overlapping sphere model.
The results are in very good agreement with data for the original
model.
\end{abstract}

%\narrowtext
\begin{multicols}{2}

\begin{picture}(0,0)
\put (37,236){\parbox{16.0cm}{\small \raggedleft %
{\em Appeared in} PHYSICAL REVIEW E, VOL. 56 PAGES 3203--3212, SEPTEMBER 1997}}
\end{picture}
\begin{picture}(0,0)
\put (-5,-486){\parbox{18.0cm}{\small \raggedright %
* Current address: University of Oxford, Department of Materials,
Parks Rd, Oxford OX1 3PH, England}}
\end{picture}
Predicting the macroscopic properties of composite
or porous materials with random microstructures is
an important problem in a range of
fields~\cite{TorqRev91,Sahimi93}.
There now exist large-scale 
computational methods for calculating the properties
of composites  given a
digital representation of their microstructure
(eg.\ permeability~\cite{Bentz94a,Adler90},
conductivity~\cite{Bentz94a,Adler90,Roberts95a}
and elastic moduli~\cite{Garboczi95a}). 
A critical problem is actually obtaining an accurate
three-dimensional description of this
microstructure~\cite{Bentz94a,Crossley91,Yao93}.
For particular materials it may be possible to simulate
microstructure formation from first principles.
Generally this relies on a detailed knowledge of the
physics and chemistry of the system; the accurate
modelling of each material requiring a significant amount
of research. Where such information is unavailable an alternative
is to {\em directly}%
~\cite{Odgaard90,Kwiecien90,MacDonald95,%
Fredrich95,Spanne94,Schwartz94,Rintoul96a}
or {\em statistically}%
~\cite{Bentz94a,Adler90,Yao93,Joshi74,Quiblier84,Adler92,%
Ioannidis95,Giona96,Losic97}
reconstruct the
microstructure from experimental images.

Several techniques of {\em direct} reconstruction
have been implemented.
A composite can be repeatedly sectioned, imaged and
the results combined to reproduce a three-dimensional
digital image of the
microstructure~\cite{Odgaard90,Kwiecien90,MacDonald95}.
For porous materials, time consuming sectioning can been
avoided by using laser microscopy~\cite{Fredrich95}
which can image pores to depths of around 150$\mu$m.
Recent micro-tomography
studies have also directly imaged the three-dimensional
microstructure of porous sandstones~\cite{Spanne94,Schwartz94} and
magnetic gels~\cite{Rintoul96a}. 
The complexity and restrictions of these methods
provide the impetus to study alternative reconstruction
methods.

Based on the work of Joshi~\cite{Joshi74}, Quiblier~\cite{Quiblier84}
introduced a method of generating a three-dimensional
{\em statistical} reconstruction of a random composite.
The method is based on matching statistical
properties of a three-dimensional model to those
of a real microstructure.
A key advantage of this approach is that the required information
can be obtained from a two dimensional image of the sample.
Recently the method has been applied to the
reconstruction of sandstone~\cite{Adler90,Yao93,Adler92,Ioannidis95}
and a material composed of overlapping spheres~\cite{Bentz94a}.
Computations of the permeability and
conductivity~\cite{Bentz94a,Adler90,Adler92} of the
reconstructed images underestimate experimental data
by around a factor of three.
This can be partially attributed to the fact that
percolation threshold of the reconstructed models
is around 10\% while the experimental systems had thresholds
of less than 3\%~\cite{Bentz94a}. Recent work in
microstructure modelling has led to a general
scheme~\cite{Roberts95a,Roberts96a,Roberts95b%
,Knackstedt95a,Roberts96b,Roberts96c,RobertsAero}
(\S~\ref{newmodels})
which includes the model employed by Quiblier.
Importantly, other models in the scheme can mimic
the low percolation thresholds observed in sandstones
(and many other materials~\cite{Roberts96a}).
It is therefore timely to reconsider statistical
methods of reconstructing composite microstructure.

Prior methods of statistical reconstruction 
produce three-dimensional models which share
first (volume fraction) and second (two-point correlation function)
order statistics with the original
sample. However the complete statistical description of
a random disordered material requires higher order 
information~\cite{Yao93,Brown55} (eg.\ the three and four point correlation
functions). Information which in turn is a crucial ingredient of
rigourous theories of macroscopic
properties~\cite{TorqRev91,Brown55,Milton81a}, and therefore important
to the success of the model.
In this paper we show that reconstructions based on
matching first and second order statistics do not necessarily
provide good models of the original composite (\S~\ref{recon1}).
An alternative
method of reconstruction is proposed and tested
(\S~\ref{recon2}).
The procedure is employed to reconstruct a composite generated from
identical overlapping spheres (IOS) and successfully predicts
the electrical conductivity of the model (\S~\ref{reconIOS}).

\section{Model composite materials}
\label{newmodels}
To study the statistical properties of composites it is
conventional to introduce a phase function
$\phi(\mbox{\boldmath $r$})$ 
which equals unity or zero as $\mbox{\boldmath $r$}$ is in phase one
or two. The volume fraction of phase one is $p=\langle \phi \rangle$,
while the standard two-point correlation function is defined as
$p^{(2)}(r)=\langle \phi(\mbox{\boldmath $r$}_1)
\phi(\mbox{\boldmath $r$}_2) \rangle$
with $r$=$|\mbox{\boldmath $r$}_2-\mbox{\boldmath $r$}_1|$ (assuming
the material is statistically homogeneous and isotropic). 
$p^{(2)}(r)$ represents the probability that two points
a distance $r$ apart will lie in phase one. From the definition
$p^{(2)}(0)=p$ and $\lim_{r\to\infty} p^{(2)}(r)\to p^2$. The 
surface area per unit volume is $s=-4 d p^{(2)}/dr|_{r=0}$~\cite{Debye57}.
Higher order functions can be analogously defined, these playing a
central role in rigourous theories of composite properties~\cite{Brown55}.
In practice the correlation functions of real composites beyond
second order are difficult to measure and there are significant
advantages in developing models for which the functions
are exactly known. The
primary models in this class are the identical overlapping sphere
model (IOS)~\cite{Weissberg63}, its generalisation to
overlapping annuli~\cite{Roberts96a}
and models derived from Gaussian random
fields (GRFs)~\cite{Roberts95a,Roberts96a,Teubner91,Berk91}
which are central to reconstruction procedures.

We utilise two methods of generating isotropic GRFs.
Each has specific advantages which we discuss. The first
method develops the random field in a cube of side-length $T$ using a
Fourier summation;
\begin{equation}  \label{yF}
y (\mbox{\boldmath$r$})=
\sum_{l=-N}^{N}
\sum_{m=-N}^{N}
\sum_{n=-N}^{N} c_{lmn} e^{i\mbox{\boldmath$k$}_{lmn}.\mbox{\boldmath$r$}}
\end{equation}
where
$\mbox{\boldmath$k$}_{lmn}=\frac{2\pi}{T}
(l\mbox{\boldmath$i$}+m\mbox{\boldmath$j$}+n\mbox{\boldmath$k$})$. 
The statistics of the field are determined by the random
variables $c_{l,m,n}$=$a_{lmn}+ib_{lmn}$ ($a_{lmn}$ and $b_{lmn}$ real).
We require that $y$ is real ($c_{l,m,n}$=$\bar c_{-l,-m,-n}$)
and that $\langle y\rangle=0$ ($c_{0,0,0}$=$0$). To ensure isotropy we 
also take $c_{lmn}$=$0$ for
$k_{lmn}$=$|\mbox{\boldmath$k$}_{lmn}|$$\ge$$2\pi N/T$.
To generate a Gaussian field the coefficients $a_{lmn}$ are taken
as random independent variables (subject to the conditions on
$c_{lmn}$) with Gaussian
distributions such that $\langle a_{lmn}\rangle=0$ and
$\langle a_{lmn}^2
\rangle$=$\frac12\rho(k_{lmn})\left({\frac{2\pi}{T}}\right)^3$
(similarly for $b_{lmn}$).
The function $\rho(k)$ is a spectral density. It can be shown
that a random field defined in this manner has field-field
correlation function
\begin{equation}
g(r)\equiv \langle y({\bf r}_1)y({\bf r}_2)\rangle
=\int_0^{\infty}{4\pi k^2 \rho (k) \frac{\sin kr}{kr}}{dk}.
\label{grho}
\end{equation}  
By convention 
$g(0)=1$ which sets a constant of proportionality on $\rho(k)$.
The definition~(\ref{yF}) can be efficiently evaluated by
an FFT routine~\cite{Roberts95a} and is $T$-periodic in
each direction. This is
valuable for approximating an infinite medium in calculations
of macroscopic properties.

Alternatively a random field can be generated using the
``random-wave'' form~\cite{Berk87,Teubner91}
\begin{equation}\label{yW}
y(\mbox{\boldmath $r$})=\sqrt{\frac{2}{N}}\sum_{i=1}^{N}
\cos(k_i \hat{\mbox{\boldmath $k$}}_i \cdot {\mbox{\boldmath $r$}} + \phi_i),
\end{equation}
where $\phi_i$ is a uniform deviate on $[0,2\pi)$ and
$\hat{\mbox{\boldmath $k$}}_i$ is uniformly distributed on a unit sphere.
The magnitude of the wave vectors $k_i$ are distributed on $[0,\infty)$ with a
probability (spectral) density $P(k)$ ($\int_0^\infty P(k)dk=1$). In terms
of the first definition $P(k)=4\pi k^2\rho(k)$.
In this case the fields are not periodic, but
$N$ can be chosen arbitrarily largely over a specified $k$ range. This
is especially useful for resolving $\rho(k)$ (so that Eq.~(\ref{grho}) holds)
in cases where it is strongly spiked (eg.\ $P(k)=\delta(k)$)~\cite{Berk91}.

Model composite materials can be defined from a GRF
$y({\mbox{\boldmath $r$}})$ by taking the region in space where
$\alpha \leq y(\mbox{\boldmath $r$}) \leq \beta$ as phase one and the
the remaining regions ($y(\mbox{\boldmath $r$}) < \alpha$ and
$y(\mbox{\boldmath $r$}) > \beta$) as phase two. This is the
``two-level cut'' random field of Berk~\cite{Berk87}. In the
case $\alpha=-\infty$ the
more common ``one-level cut'' field is
recovered~\cite{Roberts95a,Quiblier84,Teubner91}.
The phase function of this model is
$\phi(y(\mbox{\boldmath $r$}))=
H(y(\mbox{\boldmath $r$})-\alpha)-H(y(\mbox{\boldmath $r$})-\beta)$
where $H$ is the Heaviside step function.
The joint probability distribution of the correlated
random variables
$\mbox{\boldmath $y$}=
[y(\mbox{\boldmath $r$}_1),y(\mbox{\boldmath $r$}_2),\dots,
y(\mbox{\boldmath $r$}_n)]^T$ is,
$P_n(\mbox{\boldmath $y$})=( (2\pi)^ n |G|)^{-\frac12}
\exp( -\frac 12 \mbox{\boldmath $y$}^T G^{-1} \mbox{\boldmath $y$})$
where the elements of $G$ are $g_{ij}=g(r_{ij})=
\langle y({\mathbf r}_i)y({\mathbf r}_j)\rangle$.
Therefore the $n$-point correlation function is
\begin{equation}
p^{(n)}=
\int_{-\infty}^\infty\int_{-\infty}^\infty \dots \int_{-\infty}^\infty
P_n(\mbox{\boldmath $y$})
\prod_{i=1}^{n} \phi(y(\mbox{\boldmath $r$}_i))
d \mbox{\boldmath $y$}.\end{equation}
The volume fraction of phase one is $p=p^{(1)}=h=(p_\beta-p_\alpha)$  where
\mbox{$p_\alpha=(2\pi)^{-\frac12}\int_{-\infty}^\alpha e^{-t^2/2} dt$}
and $p^{(2)}(r)=h(r)$ with~\cite{Teubner91,Berk91}
\begin{eqnarray}
h(r)= && h^2+\frac{1}{2\pi}\int_0^{g(r)}
 \frac{dt}{\sqrt{1-t^2}} \times  \left[
\exp\left({-\frac{\alpha^2}{1+t}}\right) \right.  
\label{h2}
\\ && \left. \nonumber
-2\exp\left({-\frac{\alpha^2-2\alpha\beta
t+\beta^2}{2(1-t^2)}}\right)
+\exp\left({-\frac{\beta^2}{1+t}}\right) \right].
\end{eqnarray}    
The auxiliary variables $h$ and $h(r)$ are needed below.
The three-point correlation functions~\cite{Brown55} have also been
evaluated~\cite{Roberts95a,Roberts96a}.

We now show how new models can be developed.
Suppose $\phi_1(\mbox{\boldmath $r$})$ and $\phi_2(\mbox{\boldmath $r$})$
are the phase functions of
two statistically independent composites with volume fractions
$p_1$ and $p_2$ and correlation functions $p_1^{(2)}$ and $p_2^{(2)}$.
New model composites can be formed from the intersection and union
sets of each structure. The
intersection set $\phi(\mbox{\boldmath $r$})=
\phi_1(\mbox{\boldmath $r$})\!\times
\!\phi_2(\mbox{\boldmath $r$})$ has volume
fraction
$p\!=\!\langle\phi_1(\mbox{\boldmath $r$})\phi_2(\mbox{\boldmath $r$})\rangle\!=\!
\langle\phi_1(\mbox{\boldmath $r$})\rangle \langle\phi_2(\mbox{\boldmath $r$})\rangle\!=\!p_1p_2$ and correlation
function
\begin{eqnarray}
p^{(2)}(r)&=&\langle\phi_1(\mbox{\boldmath $r$}_1)\phi_2(\mbox{\boldmath $r$}_1)\phi_1(\mbox{\boldmath $r$}_2)\phi_2(\mbox{\boldmath $r$}_2)\rangle \\
          &=&
\langle\phi_1(\mbox{\boldmath $r$}_1)\phi_1(\mbox{\boldmath $r$}_2)\rangle\langle\phi_2(\mbox{\boldmath $r$}_1)\phi_2(\mbox{\boldmath $r$}_2)\rangle=
p^{(2)}_1(r)p^{(2)}_2(r). \nonumber
\end{eqnarray}
In a similar way a composite can be modelled as the union of two
independent models. In this case the phase function is
$\phi(\mbox{\boldmath $r$})=\phi_1(\mbox{\boldmath $r$})+\phi_2(\mbox{\boldmath $r$})-\phi_1(\mbox{\boldmath $r$})\phi_2(\mbox{\boldmath $r$})$ so that
$p=p_1+p_2-p_1p_2$ and
\begin{eqnarray}
p^{(2)}(r)&=&p^{(2)}_1(r)(1-2p_2)+p^{(2)}_2(r)(1-2p_1) \nonumber \\ &&+
2p_1p_2+p^{(2)}_1(r)p^{(2)}_2(r).
\end{eqnarray}
Therefore if the statistical properties of the original
morphologies are known (eg.\ level-cut GRF's or the overlapping
sphere model) the properties of their union and intersection
sets are also known~\cite{RobertsAero}.
Note that these results apply to arbitrary independent
phase functions, are simply extended to three or more independent sets,
as well as to the calculation of higher order correlation functions.
These simple results greatly extend the classes of morphology
which can be reproduced by the models.

To simplify matters we now restrict attention to
a few primary models of microstructure.
Consider first structures derived using the normal two-level cut 
GRF scheme (model ``N''). These have the basic statistical properties
$p$=$h$ (recall $h$=$p_\beta-p_\alpha$), $p^{(2)}(r)=h(r)$ and $s=-4 h'(0)$.
We also take $p_\alpha= c \times(1-p)/2$ and $p_\beta=p_\alpha+p$
($c\in[0,1]$) to
specify the level-cut parameters; for example $c=0$ corresponds to
a one-cut field ($p_{\alpha,\beta}=0,p$ or $\alpha=-\infty$) and
$c=1$ to a symmetric two-cut field ($p_{\alpha,\beta}=\frac12-\frac p2,
\frac12+\frac p2$ or $\alpha=-\beta$). 
Second, we take a class of models based on the
intersection set (model ``I'') of two statistically identical
level-cut GRF's.
For this model $p=h^2$, $p^{(2)}(r)=h^2(r)$ and
$s=-8\sqrt{p} h'(0)$ with $p_\alpha=c(1-\protect\sqrt{p})/2$ and
$p_\beta=p_\alpha+\protect\sqrt{p}$.
Finally, we introduce a model based on the union set (model ``U'')
of two two level-cut fields. In this case 
$p=2h-h^2$,
$p^{(2)}(r)=2h^2+2h(r)(1-2h)+h^2(r)$ and $s=-8\sqrt{1-p} h'(0)$ with
$p_\alpha=c\protect\sqrt{1-p}/2$ and $p_\beta=p_\alpha+1-
\protect\sqrt{1-p}$.

\begin{figure}[bt!]
{\samepage\columnwidth20.5pc
\centering \epsfxsize=8.3cm\epsfbox{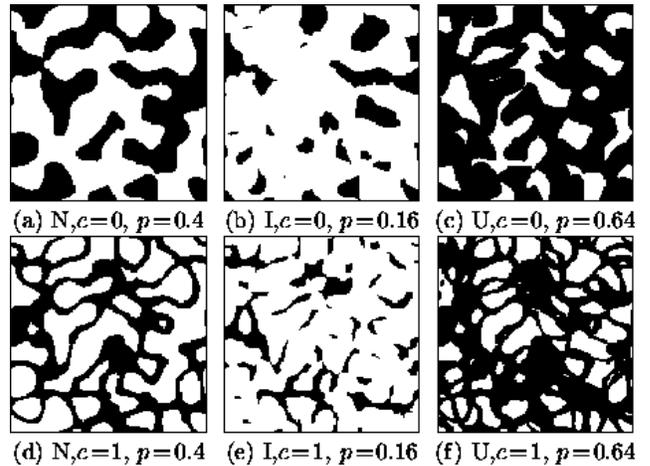}

\vskip 1mm
\caption{Six different microstructures generated by
the level-cut scheme. In the top row we show a one-cut field and
its intersection and union with a statistically identical structure.
In the bottom row we show analogous structures derived from a
two-cut field.
The images have a side length of 10$\mu$m.
\label{cutV122}}
}
\end{figure}

\begin{figure}[t!]
{\samepage\columnwidth20.5pc
\centering \epsfxsize=8.0cm\epsfbox{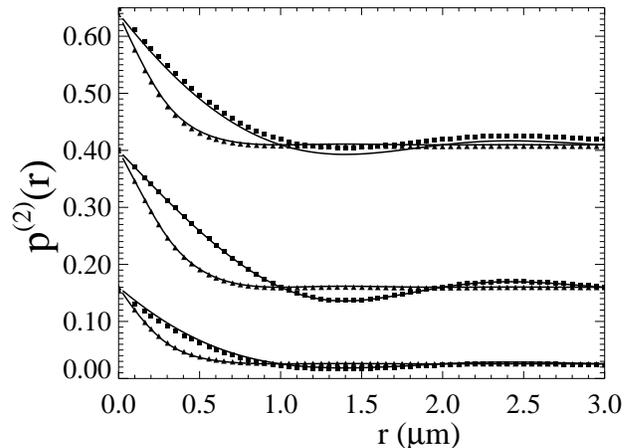}
\caption{The theoretical (lines) and measured (symbols) correlation
functions of the six models shown in
Fig.~\protect\ref{cutV122}. The squares correspond to the
models constructed from one-cut fields
(Figs.~\protect\ref{cutV122}a-c), and the triangles to the
the two-cut fields (Figs.~\protect\ref{cutV122}d-f). 
\label{p2V122}} 
}
\end{figure}

To generate examples of the models defined above we employ
the field-field correlation
function~\cite{RobertsAero,Marcelja90,Teubner87} 
\begin{equation}
g(r)=\frac{e^{-r/\xi}-(r_c/\xi)e^{-r/r_c}}{1-(r_c/\xi)}
\frac{\sin 2\pi r /d}{2\pi r /d}
\label{modelg}
\end{equation}
characterised by a correlation length $\xi$, domain scale $d$ and a cut-off
scale $r_c$. This has Fourier transform
\begin{eqnarray}
\label{modelrho}
\rho(k)&=&
\frac{\pi^{-2}(\xi-r_c)^{-1}\xi^4d^4}
{[d^2+\xi^2(kd-2\pi)^2] [d^2+\xi^2(kd+2\pi)^2]}  \\
&&-
\frac{\pi^{-2}(\xi-r_c)^{-1}r_c^4d^4}
{[d^2+r_c^2(kd-2\pi)^2] [d^2+r_c^2(kd+2\pi)^2]}. \nonumber
\end{eqnarray}
Note that $g(r)$ is symmetric in $r_c$ and $\xi$ and remains well defined
in the limits $r_c\to\xi$ and $r_c$ or $\xi\to\infty$. In the latter
cases $\rho(k)\to \delta(k-2\pi/d)/4\pi k^2$~\cite{Berk91}.
For the purposes of calculating the surface area,
\begin{equation}
-h'(0) = \frac{\sqrt{2}}{2\pi}\left( e^{-\frac12 {\alpha^2}}+
e^{-\frac12 \beta^2} \right)\sqrt{\frac{4\pi^2}{6d^2}+\frac{1}{2r_c\xi}}.
\end{equation}
In the case $r_c$ or $\xi\to0$ a fractal surface
results~\cite{Roberts96b,Berk91}.
Cross-sections of six of the model microstructures obtained
with $r_c$=1, $\xi$=2 and $d$=2$\mu$m are illustrated
in Fig.~\ref{cutV122}. $p^{(2)}(r)$ is measured from three-dimensional
realisations (using $128^3$ pixels) of the models and plotted
against its theoretical value in Fig.~\ref{p2V122}.
The agreement is very good. In the following 
section we also consider each of the models at an intermediate
value of $c=\frac12$. The extra three models, along with the six
shown in Fig.~\ref{cutV122}

\vspace{-4mm}

\vspace{-2mm}

\noindent
give nine primary classes of
microstructure
with which to compare real composites.
These broadly cover the types of morphology obtainable by
combining two composites generated by the level-cut GRF scheme.

\vspace{-6mm}

\section{Statistical reconstruction}
\label{recon1}

\vspace{-4mm}

The two most common experimentally measured morphological
quantities of composites are the volume fraction $p_{exp}$ and
the two-point correlation function
$p_{exp}^{(2)}(r)$
(eg.\ Refs.~\cite{Adler90,Ioannidis95,Losic97,Berryman85d,Blair96}).
Consider how this information might be used to
reconstruct the composite using the simple one-cut GRF model
(model N, $c\!=\!0$ or $\alpha\!=\!-\infty$). The level-cut parameter
$\beta$ can be obtained by solving
\mbox{$p_{exp}=(2\pi)^{-\frac12}\int_{-\infty}^\beta e^{-t^2/2} dt$}
and the field-field function obtained by numerical inversion
of 
\begin{equation}
\label{invone}
p_{exp}^{(2)}(r)=p_{exp}^2+\frac{1}{2\pi}\int_0^{g(r)}
 \frac{dt}{\sqrt{1-t^2}} \exp\left({-\frac{\beta^2}{1+t}}\right).
\end{equation}
From $g(r)$ we can obtain $\rho(k)$ by inverting Eq.~(\ref{grho})
and using either Eq.~(\ref{yF}) or~(\ref{yW}) to obtain $y(\mbox{\boldmath $r$})$
and hence the model phase function $\phi(\mbox{\boldmath $r$})$.
The reconstruction shares first and second order
statistical properties with the image and would
therefore be expected to yield
a reasonable model of the original composite.
This is similar to the procedure of Quiblier~\cite{Quiblier84}
employed in previous
studies~\cite{Bentz94a,Adler90,Yao93,Adler92,%
Ioannidis95,Giona96,Losic97}
although the formulation of the model is different.
There are several operational problems
with this reconstruction procedure.
First, the numerical inversion of Eq.~(\ref{invone})
may not be robust or well defined.
Furthermore experimental error in $p^{(2)}_{exp}(r)$ is carried over
to $g(r)$. Second, the inversion of Eq.~(\ref{invone})
may yield a spectral density $\rho(k)$ which is not strictly positive.
We now generalise the method to incorporate the models N, I and U of
\S~\ref{newmodels} and show how these problems
can be avoided.

First select one of the three models (N, I or U) and a value
of $c$=$0$, $\frac12$ or $1$ (giving a total of nine
combinations) so that $\alpha$ and $\beta$ are fixed
by $p_{exp}$. It remains to find $g(r)$. Instead of inverting an
analog of Eq.~(\ref{invone}) we assume this function is of the general 
form given by Eq.~(\ref{modelg})
(this guarantees that $\rho(k)$ is positive).
The three length scale parameters
are obtained by a best fit procedure which minimises the
normalised least-squares error;
\begin{equation}
Ep^{(2)}=\sum_{i=1}^{M} [p^{(2)}_{fit}(r_i)-p^{(2)}_{exp}(r_i)]^2
/ \sum_{i=1}^{M} [p^{(2)}_{exp}(r_i)-p_{exp}^2]^2.
\label{Ep2}
\end{equation}
Here $p^{(2)}_{fit}(r_i)=p^{(2)}[g(r_i;r_c,\xi,d)]$ is the 
correlation function appropriate for model N, I or U.
Once $r_c$, $\xi$ and $d$ have been obtained the reconstruction
$\phi(\mbox{\boldmath $r$})$ can be generated.
If the one-cut model (N, $c\!=\!0$) is chosen we
assume that the results will not differ significantly 
from those obtained using Quiblier's method.

\begin{figure}[h!]
{\samepage\columnwidth20.5pc
\centering \epsfxsize=8.0cm\epsfbox{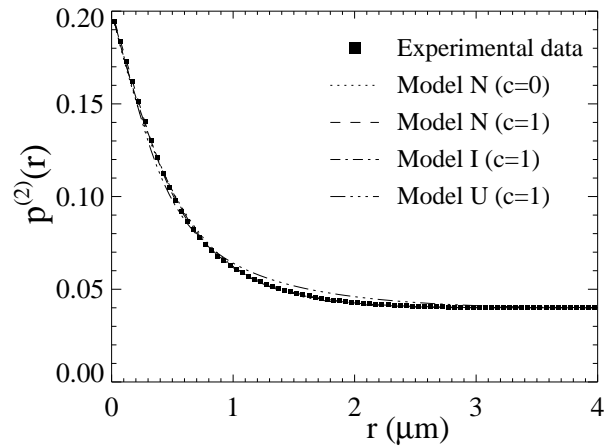}
\caption{The correlation functions $p_{fit}^{(2)}(r)$ (lines) of four
reconstructed models obtained by fitting
``experimental'' data (symbols).
\label{p2II}}
}
\end{figure}

To illustrate the procedure we reconstruct a material
with known statistical properties. For this purpose we
choose a normal two-cut GRF model with $p_{\alpha,\beta}=0.4,0.6$
(ie.\ model N, $c=1$) obtained from the
field-field function~\cite{Roberts95a}
\begin{equation}
g(r)=e^{-(r/l_0)^2};\;\;\;
\rho(k)=\frac{l_0^3}{(4\pi)^\frac32}e^{-(k l_0/2)^2}
\end{equation}
with $l_0=2.0\mu$m. The ``experimental'' data for the reconstruction
$p^{(2)}_{exp}(r_i)$ are evaluated using Eq.~(\ref{h2})
at $80$ points distributed uniformly on the interval
$[0,4]\mu$m (shown as symbols in Fig.~\ref{p2II}).
The minimisation algorithm is used to find
$r_c$, $\xi$ and $d$ for four different models. Numerical results
are reported in Table~\ref{tabIIa} and the best-fit functions
$p^{(2)}_{fit}$ are plotted in Fig.~\ref{p2II}.
Each of the models is able to provide an excellent fit of the data.
As expected, model N ($c=1$) provides
the least value of $Ep^{(2)}$. However the

\noindent
\begin{minipage}[t!]{8.5cm}
\begin{table}
\caption{The parameters obtained in the reconstruction procedure
(Eq.~(\protect\ref{Ep2})) of a test composite.  The
surface area of the original model is $0.87\mu$m$^{-1}$.
Here, and in subsequent tables, $n(m)$ denotes $n\!\times\!10^{-m}$.
\label{tabIIa}}
\begin{tabular}{ccccccc}       
Cl&$c$&$r_c$&$\xi$&$d$&$Ep^{(2)}$&$s_{fit}$ \\
\hline
N&0&0.4033&0.4031&7.7069&1(-3)&1.13\\
N&1&2.3702&2.3688&6.2140&3(-5)&0.89\\
I&1&0.9739&0.9729&9.1032&4(-4)&1.05\\
U&1&4171.1&6651.8&8.3899&4(-3)&0.98\\
\end{tabular} 
\end{table}
\end{minipage}

\noindent

\begin{figure}[bt!]
{\samepage\columnwidth20.5pc
\centering \epsfxsize=8.3cm\epsfbox{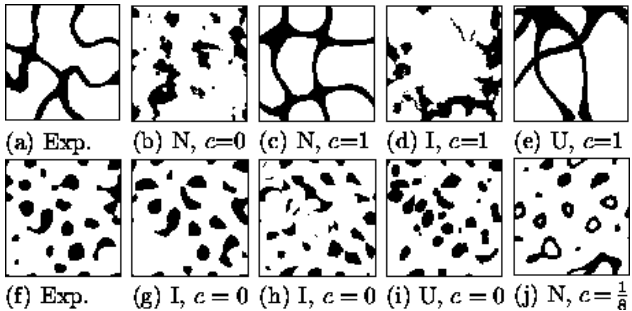}

\vspace{1mm}

\caption{Realisations of ``experimental'' and reconstructed
composites. Top row: A material with a monotonicly decaying
correlation function (a) compared with four reconstructions (b)-(e).
The two point correlation functions of each composite are
practically identical (see Fig.~\ref{p2II}).
Bottom row: A model composite exhibiting an oscillatory
correlation function (f) and four reconstructions (g)-(j).
In each case the region shown is 10$\times$10$\mu$m.
\label{xsecII}}
}
\end{figure}

\vspace{-2mm}

\noindent
relative improvement
over the other three models is
not large, and probably of little significance in the presence of
experimental error.
Cross-sections of the original composite and the reconstructions
are shown in Fig.~\ref{xsecII}(a)-(e). The extremely different
morphologies
exhibited by the reconstructions provide a graphical illustration
of the non-uniqueness of $p^{(2)}(r)$.
Therefore for prediction of macroscopic
properties (which will differ dramatically for materials shown
in Fig.~\ref{xsecII}) it is necessary to find a more discriminating
method of distinguishing composites.
From the cross-sectional images the best candidates appear to be
models N ($c\!=\!1$) and U ($c\!=\!1$) shown 
in Figs.~\ref{xsecII}(c) and (e). Obviously it is preferable to establish
some quantitative test to choose the best representation. 

\begin{figure}[bt!]
{\samepage\columnwidth20.5pc
\centering \epsfxsize=8.0cm\epsfbox{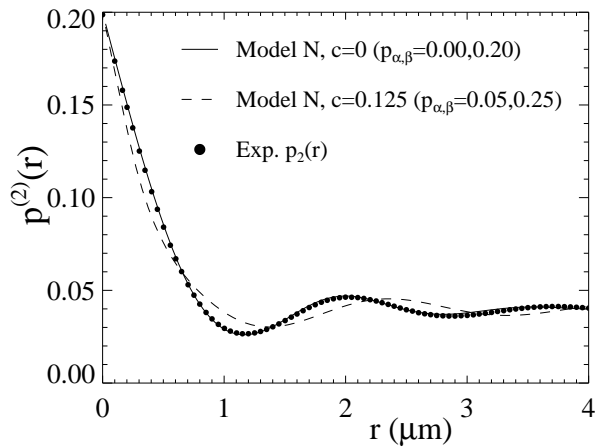}
\caption{Correlation functions of two reconstructions (lines) of a
material exhibiting an oscillatory $p^{(2)}(r)$ (symbols).
A ``mild'' two-cut model (dashed line) is unable to 
accurately reproduce the strong oscillations.
 \label{p2III}} 
}
\end{figure}

A second useful illustration of the method is provided by
reconstructing a material with a strongly oscillating
correlation function.
For this case we take as a test-composite
a one-cut model with $p=0.2$ and
$p_{\alpha,\beta}=0.0,0.2$ (ie.\ model N, $c=0$)  based on
the field-field
 function~\cite{Roberts95a}
\begin{eqnarray}
\nonumber
g(r)&=&3r^{-3}(k_1^3-k_0^3)^{-1} (\sin k_1r - \sin k_0r) \\
    & &-  3r^{-2}(k_1^3-k_0^3)^{-1}(k_1\cos k_1r - k_0\cos k_0r ) \\
\rho(k)&=&3[4\pi(k_1^3-k_0^3)]^{-1}[H(k-k_0)-H(k-k_1)]
\end{eqnarray}
with $k_0=3.0$ and $k_1=4.5$($\mu$m)$^{-1}$.
The oscillatory behaviour of the correlation function
(see Fig.~\ref{p2III}) can only

\noindent

\vspace{-2mm}

\vspace{-2mm}

\noindent
be reproduced by three 
of the nine basic microstructures; models N,I and U with $c=0$
(ie.\ those formed from one-cut fields).
For these models $Ep^{(2)}\!<\!0.005$ whereas $Ep^{(2)}\!>\!0.02$ for those
based on two-cut structures ($c\!\geq\!\frac12$ so
$0\!<\!p_\alpha\!<\!p_\beta$).
To illustrate this we show the best fit of a normal two-cut model
with $p_{\alpha,\beta}=0.05,0.25$ (N, $c\!=\!\frac18$).
As can be seen in Fig.~\ref{p2III} this ``mild'' two-cut
model (shown as a dashed line) cannot reproduce the behaviour
of the experimental data (see Table~\ref{tabIIIa}).
Realisations of the original material and reconstructions
are shown in Fig.~\ref{xsecII}(f)-(j). Each
appears to provide a reasonable representation. 

In contrast to the case of a monotonicly decaying
$p^{(2)}(r)$ (which was reproduced by four distinct models)
strong oscillations appear to be a signature of morphologies
generated by the single level-cut model. Unless there
exists some reason to employ models U and I in such a case
it is likely that the standard one-cut GRF
(ie.\ the model employed in prior studies)
will be appropriate.

\noindent
\begin{minipage}{8.5cm}
\begin{table}
\caption{Reconstruction of a normal one-cut model with an oscillatory
correlation function. Models formed from
two-cut fields (ie.\ $p_\alpha>0$) were unable
to reproduce the oscillations of $p^{(2)}(r)$ (see eg.\ row 4).
The surface area of the original model is $1.00\mu$m$^{-1}$.
\label{tabIIIa}}
\begin{tabular}{ccccccc}
Cl&$c$&$r_c$&$\xi$&$d$&$Ep^{(2)}$&$s_{fit}$ \\
\hline
N&0&1.6326&1.6330&1.6586&2(-4)&1.01\\
I&0&2.8276&2.8305&1.7220&4(-3)&1.20\\
U&0&3.9019&3.8935&1.7263&4(-3)&1.10\\
N&$\frac18$
   &4.6684&4.6893&1.9215&3(-2)&1.28
\end{tabular}
\end{table}
\end{minipage}

\noindent
There is also a physical basis for this
argument when spinodal decomposition plays a role in the
microstructural formation.
In this case Cahn~\cite{Cahn65} has shown that
the evolution of the phase interface is described by the 
level-set of a sum of random waves similar to (\ref{yW}).

\begin{figure}[bt!]
{\samepage\columnwidth20.5pc
\centering \epsfxsize=8.0cm\epsfbox{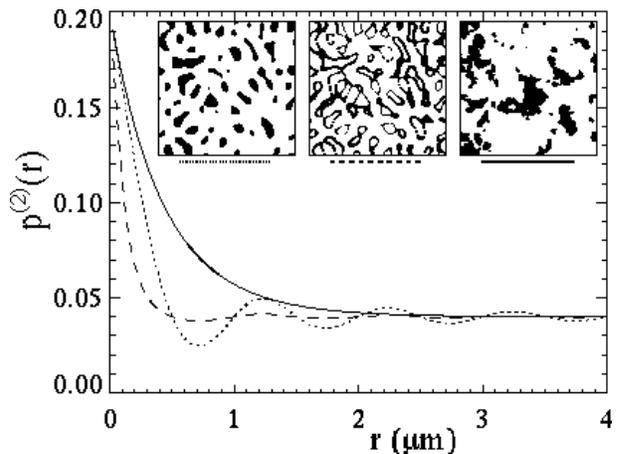}
\caption{Three different types of microstructure.
A one-cut model with a well defined domain (or repeat) scale
(left), a two-cut model obtained from the same GRF (centre) and
a one-cut field with no domain scale (right). The oscillations
of $p^{(2)}$ are very weak for the central model even though
the domain scale is obvious to the eye.
\label{compile}}
}
\end{figure}

Finally we comment on the morphological origin of
the oscillations, and why they cannot be well
reproduced by two-cut models. In Fig.~\ref{compile}
we show $p^{(2)}(r)$ and an image of model N, $c=0$
with $r_c$=2, $\xi$=4 and $d$=1$\mu$m. The material
has strong oscillatory correlations, these representing
the ``regular'' alternating domains which appear in the
image. Compare this with data shown for the
two-cut model (N, $c\!=\!\frac12$) obtained from the same GRF:
the alternating structure is still present but
the oscillations are practically extinguished. This is due
to the sharper decay (or equivalently the doubled specific surface)
associated with the thinner two-cut structures~\cite{RobertsAero}.
For comparison we also show a structure with no repeat
scale (model N, $c=0$ with $r_c$=$\frac16$, $\xi$=$\frac12$ and
$d$=100$\mu$m).

\noindent

\section{Comparison of higher-order statistical properties}
\label{recon2}

We have shown that reconstructions
exhibiting quite different morphological properties can share
the same two-point correlation function. Here we
propose and test three methods with the aim of finding a way
of selecting the best reconstruction.
Following Yao {\em et al}~\cite{Yao93} we can compare the three-point
correlation function of the model and experimental materials.
To do so we define a normalised least-square measure of the error as
\begin{eqnarray}
\nonumber
Ep^{(3)}&=&
\sum_{i=1}^{N_r}
\sum_{j=1}^{N_s}
\sum_{k=1}^{N_\theta}
[p^{(3)}_{fit}(r_i,s_j,\theta_k)-p^{(3)}_{exp}(r_i,s_j,\theta_k)]^2 \\
&&\div 
\sum_{i=1}^{N_r}
\sum_{j=1}^{N_s}
\sum_{k=1}^{N_\theta}
[p^{(3)}_{exp}(r_i,s_j,\theta_k)-p_{exp}^3]^2.
\label{Ep3}
\end{eqnarray}
The three-point function $p^{(3)}(r,s,\theta)$ gives the
probability that three points distances $r$, $s$ and $t\!=\!(r^2+s^2-
2rs\cos\theta)^\frac12$ apart all lie in phase one. For our examples
we take $N_{r,s,\theta}=8$ with a uniform distribution of $r$ and $s$
on $[0,2]\mu$m and $\theta$ on $[0,\pi]$

A second method of characterising morphology is to calculate 
microstructure parameters which
appear in theoretical bounds on transport
and elastic properties~\cite{TorqRev91,Milton81a}.
We therefore expect the parameters to contain
critical information about the aspects of microstructure relevant
to macroscopic properties. These are
\begin{eqnarray}
\label{zeta}
\zeta&=&
\frac9{2pq}\int_0^\infty\!\!\frac{dr}{r} \int_0^\infty
\!\! \frac{ds}{s} \int_{-1}^1 \!\!\! du \;  P_2(u) f(r,s,\theta) \\
\eta&=&\frac {5\zeta}{21}+
\frac{150}{7pq}\int_0^\infty\!\!\frac{dr}{r}
\int_0^\infty\!\! \frac{ds}{s}
\int_{-1}^1 \!\!\! du \; P_4(u) f(r,s,\theta)
\label{eta}
\end{eqnarray}
where $f(r,s,\theta)=p^{(3)}(r,s,\theta)-p^{(2)}(r)p^{(2)}(s)/p$, $q=1-p$,
$u=\cos\theta$ and $P_n(u)$ denotes the Legendre polynomial of order $n$.
The parameter $\zeta$ occurs in bounds on the conductivity and
the bulk modulus, while $\eta$ occurs in bounds
on the shear moduli. As $p^{(3)}_{fit}$ and $p^{(3)}_{exp}$ are 
available for our test models the parameters can be 
calculated~\cite{Roberts95a,Roberts96a}. Techniques have also been
suggested for directly evaluating the parameters from
experimental images~\cite{Coker95,Helsing95}. We anticipate that
the closer $\zeta_{fit}$ is to $\zeta_{exp}$ the better the reconstructed
model. Note that $\zeta$ and $\eta$ contain only third order
statistical information and higher order information is
potentially important for our purposes.

A third simple measure of microstructure is the chord-distribution
function of each phase~\cite{Coker95,Levitz92,TorqLu93}.
For phase one this is obtained by placing 
lines through the composite and counting the number of chords $n(r)$
of a given length $r$ which lie in phase one. The chord-distribution
is defined as $\rho^{(1)}(r)=n(r)/\int_0^\infty n(r) dr$ so that
$\rho^{(1)}(r)dr$ is the probability that a randomly selected chord
will have length between $r$ and $r+dr$. $\rho^{(2)}(r)$
is defined in an analogous manner. At present it is not possible to
analytically evaluate this function for the level-cut GRF media,
but it can be simply evaluated from realisations of the experimental
and reconstructed materials. To quantify the difference between the
chord-distributions we again employ a least-squares error;
\begin{equation}
E\rho^{(j)}=\sum_{i=1}^{M} [\rho^{(j)}_{rec}(r_i)-\rho^{(j)}_{exp}(r_i)]^2
/\sum_{i=1}^{M} [\rho^{(j)}_{exp}(r_i)]^2
\end{equation}
with $j=1,2$. Note that $\rho^{(j)}(r)$ contains information about the
degree of connectedness in phase $j$ and thus is likely to incorporate
important information regarding macroscopic properties~\cite{prenote2}.

We also compute
the conductivity of samples (size 128$^3$ pixels)
using a finite-difference scheme~\cite{Roberts95a}.
We choose the conductivity of phase one as $\sigma_1=1$ (arbitrary
units) and phase two insulating ($\sigma_2=0$). At this contrast
the effective conductivity $\sigma$ is very sensitive to
microstructure. The results therefore allow us to gauge the ability
of a reconstruction to predict macroscopic properties.
This contrast also occurs commonly in a range of systems
(eg.\ electrical conductivity of brine saturated porous rocks or
thermal conductivity of aerogels and foams).

\begin{figure}[bt!]
{\samepage\columnwidth20.5pc
\centering \epsfxsize=8.0cm\epsfbox{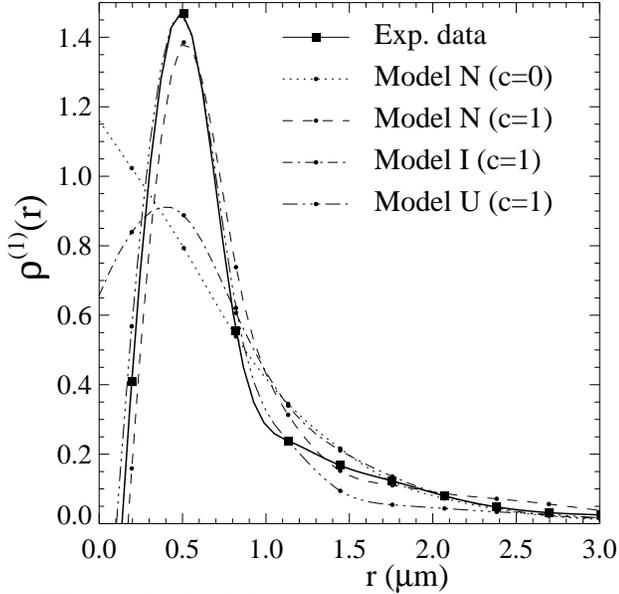}
\caption{The chord-distribution (for phase 1) of an ``experimental''
composite (Fig.~\ref{xsecII}a) compared with data for
the four reconstructions shown in Figs.~\ref{xsecII}(b)-(e). Both
models N and U ($c=1$) appear to mimic the ``experimental'' data.
The lines in the graph are guides to the eye only.
\label{cho1II}} 
}
\end{figure}

\begin{figure}[bt!]
{\samepage\columnwidth20.5pc
\centering \epsfxsize=8.0cm\epsfbox{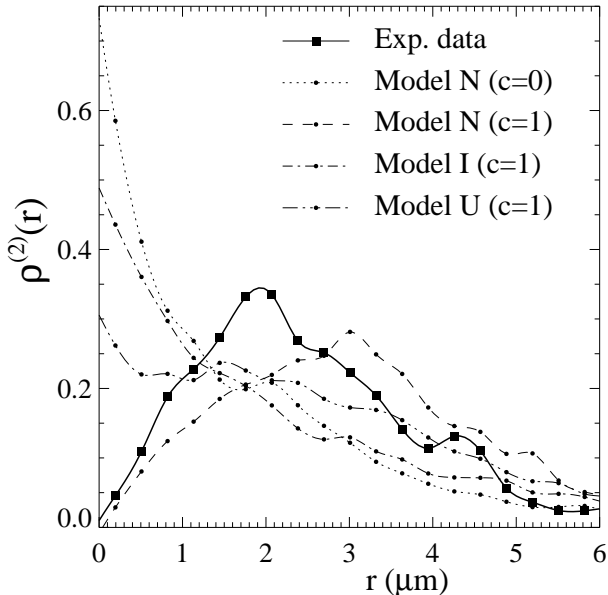}
\caption{The chord-distribution (for phase 2) of an ``experimental''
composite compared with data for four reconstructions (see caption
of Fig.~\ref{cho1II}).
\label{cho2II}} 
}
\end{figure}

We have calculated the morphological quantities defined above
for the first four reconstructions (reported in Table~\ref{tabIIa}).
The results are shown in Table~\ref{tabIIb}.
First note that $Ep^{(3)}$ is greater
than $Ep^{(2)}$ by a factor of 2-5~\cite{prenote3} in each case
and is probably of little use in an actual reconstruction. 
The values of the microstructure parameters $\zeta$ and $\eta$ are
conclusive; as we expect they indicate that model N ($c=1$) is best.
The chord-distributions of the experimental and reconstructed material
are shown in Fig.~\ref{cho1II} (phase one) and Fig.~\ref{cho2II} (phase two). 
From Table~\ref{tabIIb} we see that the chord-distribution provides a very
strong signature of microstructure. The results indicate that either
model N ($c\!=\!1$) or model U ($c\!=\!1$) is the best
reconstruction. The fact that the conductivity of each model is so 
close to the experimental data provides some evidence
that matching the chord distributions is more important than
matching $\zeta$ and $\eta$.
The same comparison is shown for the reconstructions of the test
composite which exhibits an oscillatory $p^{(2)}(r)$ in
Table~\ref{tabIIIb}.
Model N ($c\!=\!0$) provides the best reconstruction based
on both the chord-distribution and the microstructure parameters.
This leads to a good prediction of the conductivity.

\noindent
\begin{minipage}{8.5cm}
\begin{table}
\caption{A comparison of the statistical and transport properties
of the four reconstructed models (Table~\protect\ref{tabIIa})
with those of the ``experimental'' composite. The measured
surface area of the digital reconstructions is also shown.
\label{tabIIb}}
\begin{tabular}{ccccccccc}       
Cl&$c$&$Ep^{(3)}$&$\zeta_{fit}$&$\eta_{fit}$&$s_{rec}$&$E\rho^{(1)}$&
$E\rho^{(2)}$&$\sigma_{rec}/\sigma_1$ \\
\hline
N&0&5(-3)&0.32&0.29&1.06&0.25&0.62&0.032\\
N&1&9(-5)&0.74&0.54&0.75&0.04&0.11&0.114\\
I&1&2(-3)&0.47&0.37&0.98&0.20&0.48&0.069\\
U&1&6(-3)&0.87&0.70&1.02&0.02&0.15&0.120\\
\hline
\multicolumn{3}{l}{``Exp.'' data}&%
\multicolumn{1}{c}{0.72}&% 
\multicolumn{1}{c}{0.54}&% 
\multicolumn{1}{c}{0.87}&% 
\multicolumn{1}{c}{}&% 
\multicolumn{1}{c}{}&% 
\multicolumn{1}{c}{0.110} 
\end{tabular} 
\end{table}
\end{minipage}

\vskip 4mm 

\vspace{2mm}

In \S~\ref{recon1} we showed that it was possible to generate
a number of morphologically distinct reconstructions which
share first and second order statistical properties with
an experimental composite. Here we have suggested three methods
of choosing the best reconstruction. 
As $Ep^{(3)}$ is relatively small for all seven reconstructions
shown in Tables~\ref{tabIIb} and ~\ref{tabIIIb}, $p^{(3)}$ (like $p^{(2)}$)
does not appear to provide a strong signature of
microstructure~\cite{prenote3}.
It is therefore not possible to conclude that a good
reproduction of $p^{(3)}$ (or $p^{(4)}$) implies a successful
reconstruction as was done in Ref.~\cite{Yao93}.
In contrast both the chord-distributions and the microstructure
parameters appear to provide a strong signature of
composite morphology, and hence a method of selecting a useful
reconstruction of the original material.

\noindent
\begin{minipage}{8.5cm}
\begin{table}
\caption{A comparison of the statistical and transport properties
of the three reconstructed models (Table~\protect\ref{tabIIIa}) which
are able to reproduce the oscillatory correlation function of
a test composite.
\label{tabIIIb}}
\begin{tabular}{ccccccccc}       
Cl & $c$ & $Ep^{(3)}$ & $\zeta_{fit}$ & $\eta_{fit}$ &
$s_{rec}$ & $E\rho^{(1)}$ &
$E\rho^{(2)}$ & $\sigma_{rec}/\sigma_1$ \\
\hline
N&0&9(-5)&0.24&0.20&1.00&0.001&0.003&0.025\\
I&0&5(-3)&0.33&0.25&1.16&0.137&0.036&0.032\\
U&0&5(-3)&0.20&0.17&1.10&0.008&0.127&0.009\\
\hline
\multicolumn{3}{l}{``Exp.'' data} & 
\multicolumn{1}{c}{0.24\tablenotemark[1]} & 
\multicolumn{1}{c}{0.20\tablenotemark[1]} & 
\multicolumn{1}{c}{} & 
\multicolumn{1}{c}{} & 
\multicolumn{1}{c}{} & 
\multicolumn{1}{c}{0.023} 
\end{tabular} 
\tablenotetext[1]{Ref.~\cite{Roberts95a}}
\end{table}
\end{minipage}

\section{Reconstruction of the IOS model}
\label{reconIOS}
Realisations of the identical overlapping
sphere (IOS) model~\cite{Weissberg63}
(or Poisson grain model~\cite{Andraud97}) are generated by
randomly placing spheres into a solid or void.
In the latter case the morphology is thought to
provide a reasonable model of the pore-space in granular rocks
(so transport occurs in the irregular void region).
As the model has a different structure to the level-cut
GRF model it provides a useful
test of reconstruction procedures~\cite{Bentz94a}.
The correlation function of the material~\cite{Weissberg63}
is $p^{(2)}(r)=p^{v(r)}$ for $r<2r_0$ and
$p^{(2)}(r)=p^2$ for $r\geq 2r_0$ where
\begin{equation}
v(r)=1+\frac34 \left(\frac r{r_0}\right)
      -\frac1{16}\left(\frac r{r_0}\right)^3.
\end{equation}
For this model it is also possible to calculate the pore chord distribution
as $\rho^{(1)}(r)=-3/4r_0\times \ln p \; p^{3r/4r_0}$~\cite{TorqLu93}.

We first consider the IOS model at volume fraction $p_{exp}=0.2$.
The system is 80\% filled with spheres of radius $r_0$=1$\mu$m.
Nine reconstructions are generated (by minimising $Ep^{(2)}$),
and their higher order statistical properties are compared with those
of the IOS model in Table~\ref{tabIOSa}.
Based on $Ep^{(2)}$ (and $Ep^{(3)}$) we note that
models U ($c=\frac12,1$) perform poorly while the standard
one-cut model is very good.
The microstructure parameters $\zeta$ and $\eta$ indicate
that the best reconstruction is model I ($c=1$)
followed by model I ($c=\frac12$).
However both models fail to reproduce the 
solid chord distribution ($E\rho^{(2)}>0.6$) which is better mimicked
by models I ($c=0$) and N ($c=0$). The ambiguity of the results
indicate that none of models considered may be appropriate.
 
\noindent
\begin{minipage}{8.5cm}
\begin{table}[t]
\caption{A comparison of the statistical properties of 11
reconstructions with those of the IOS model at porosity 20\%.
Most of the models are able to reproduce the low order
statistical properties of the IOS model.
\label{tabIOSa}}
\begin{tabular}{clccccccc}       
Cl&$c$&$Ep^{(2)}$&$Ep^{(3)}$&$s_{fit}$&
$\zeta_{fit}$&$\eta_{fit}$&$E\rho^{(1)}$&$E\rho^{(2)}$ \\
\hline
N&0        &1(-4)&9(-4)&0.94&0.31&0.28&0.066&0.26\\
N&$\frac12$&3(-3)&5(-3)&0.79&0.74&0.54&0.35&0.15 \\
N&1        &2(-3)&8(-3)&0.79&0.84&0.63&0.59&0.31 \\
I&0        &2(-4)&7(-4)&0.98&0.35&0.30&0.024&0.24\\
I&$\frac12$&6(-4)&1(-3)&1.07&0.50&0.38&0.042&0.65\\
I&1        &4(-4)&1(-3)&1.05&0.52&0.40&0.030&0.63\\
U&0        &2(-4)&1(-3)&0.92&0.28&0.26&0.077&0.30\\
U&$\frac12$&1(-2)&2(-2)&0.91&0.79&0.62&0.49&0.11\\
U&1        &1(-2)&2(-2)&0.91&0.87&0.70&0.40&0.15\\
I$_5$&     &7(-4)&6(-4)&1.00&0.40&0.33&0.003&0.23\\
I$_{10}$&  &1(-3)&5(-4)&1.00&0.43&0.35&0.003&0.13\\
\hline
\multicolumn{4}{l}{``Exp.'' data (IOS)} & 
\multicolumn{1}{c}{0.96} & 
\multicolumn{1}{c}{0.52} & 
\multicolumn{1}{c}{0.42} & 
\multicolumn{1}{c}{    } & 
\multicolumn{1}{c}{    } 
\end{tabular} 
\end{table}
\end{minipage}

\begin{figure}[bt!]
{\samepage\columnwidth20.5pc
\centering \epsfxsize=8.0cm\epsfbox{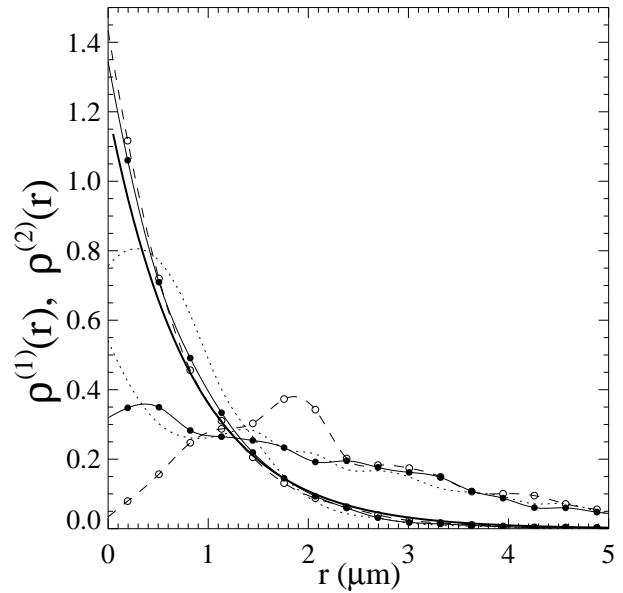}
\caption{The chord distribution of the IOS model (open symbols),
model I$_{10}$ (solid symbols) and the standard one-cut
model (broken line, symbols omitted for clarity).
The heavy line is the theoretical curve for the IOS model
and the lighter lines are guides to the eye only.
\label{choIOS}}
}
\end{figure}

The IOS model can be thought of as the intersection set of infinitely
many composites comprised of a single sphere of phase 2
(so $\phi(\mbox{\boldmath $r$})=0$ within the sphere). This suggests that
the morphology may be better modelled with the level-cut scheme by increasing
the number of primary composites beyond two.
To this end we generalise model I to the case of $n$ independent
one-cut fields so that $p^{(2)}(r)=h^n(r)$ with $p_\alpha=0$,
$p_\beta=p^{(1/n)}$ and $s=-4np^{1-1/n}h'(0)$. This is termed model
``I$_n$''. The statistical properties of the reconstructions
for the cases $n$=5 and $n$=10 are shown in rows 10 and 11 of
Table~\ref{tabIOSa}.
The models reproduce the ``experimental'' pore chord distribution
very well, and offer a progressively better representation
of the solid chord distribution and microstructure parameters.
The chord-distributions of model I$_5$ are shown in Fig.~\ref{choIOS} along
side those of the standard one-cut model and the IOS model.
The good agreement between the measured and theoretical
value of $\rho^{(1)}(r)$ for the IOS model demonstrates the
accuracy with which this function can be evaluated for a
sample of 128$^3$ pixels.

\begin{figure}[t!]
{\samepage\columnwidth20.5pc
\centering \epsfxsize=8.5cm\epsfbox{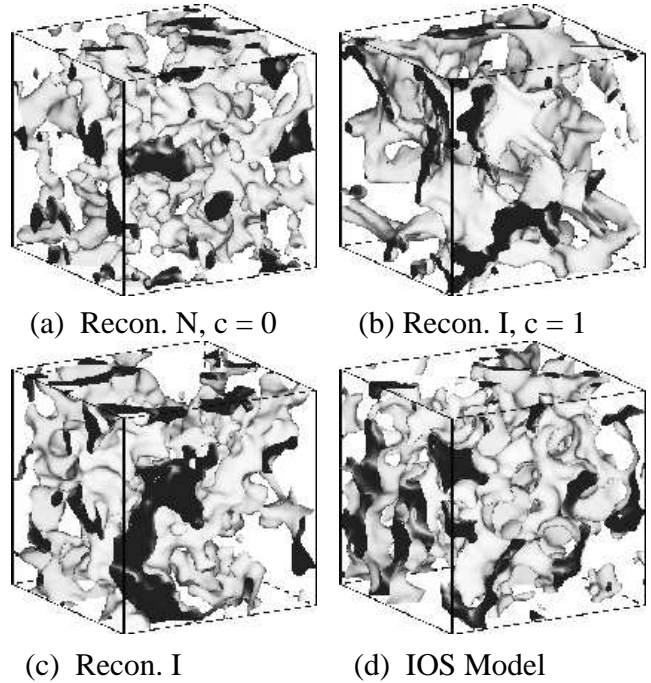}

\vskip 3mm
\caption{Reconstructions of the overlapping sphere (IOS) model at porosity
$p=0.2$. To aid visualisation the pores are shown as solid, and solid
as void. The images shown here and the chord distributions
(Fig.~\ref{choIOS}) indicate model I$_{10}$ provides the best reconstruction
of the IOS model.
\label{3DIOS}}
}
\end{figure}

To determine which morphological measure ($\zeta$ and $\eta$ or
$Ep^{(1)}$ and $Ep^{(2)}$) should be used to select the best reconstruction
we examine the model morphology and conductivity.
Three-dimensional images of models N ($c\!=\!0$), \mbox{I ($c\!=\!1$)}
and $I_{10}$ are shown alongside the IOS model in Fig.~\ref{3DIOS}.
The pore space of the single-cut GRF (Fig.~\ref{3DIOS}a)
is more disconnected than that of the IOS model, while the pores are too
large and uniform in the intersection model (Fig.~\ref{3DIOS}b).
Model I$_{10}$ (Fig.~\ref{3DIOS}c) appears better able
to reproduce the interconnected structures characteristic
of overlapping spheres.
The results for the conductivity are, $\sigma\!=\!0.038$ for model
N ($c\!=\!0$), $\sigma\!=\!0.080$ for model I ($c\!=\!1$),
$\sigma\!=\!0.052$ for model I$_{10}$ and $\sigma\!=\!0.063$ for IOS.
The fact that model I$_{10}$ better mimics IOS morphology and
conductivity than model I ($c\!=\!1$) provides evidence that
minimising $E\rho^{(j)}$ should be given more weight than
matching experimental values of $\zeta$ and $\eta$.

\clearpage

\noindent
\begin{minipage}{8.5cm}
\begin{table}
\caption{The results of the reconstruction procedure
for the IOS model. The specific surface of the
IOS model is $s=0.71,0.96,1.08,1.10\mu$m$^{-1}$ as $p$ increases.
Generally model I$_{10}$ provides a better match of the chord
distributions than model I$_{5}$. In each case $p_\beta\!=\!p^{1/n}$ for
model I$_{n}$.
\label{tabIOSb}}
\begin{tabular}{ccccccccc}       
$p$&Cl&$r_c$&$\xi$&$d$&$Ep^{(2)}$&$s_{fit}$
&$E\rho^{(1)}$&$E\rho^{(2)}$ \\
\hline
0.1 &I$_5$ &0.8770&0.8769&3.8336&3(-4)&0.69&0.011&0.33 \\
    &I$_{10}$&1.2472&1.2470&3.8608&5(-3)&0.70&0.011&0.31 \\
0.2 &I$_5$ &0.9942&0.9947&3.9055&8(-4)&1.00&0.003&0.23 \\
    &I$_{10}$&1.4173&1.4174&3.9777&1(-3)&1.00&0.003&0.13 \\
0.3 &I$_5$ &1.0974&1.0973&3.9756&1(-3)&1.14&0.003&0.23 \\
    &I$_{10}$&1.6047&1.6053&4.0375&1(-3)&1.13&0.003&0.19 \\
0.4 &I$_5$ &1.2148&1.2151&4.0250&1(-3)&1.17&0.006&0.16 \\
    &I$_{10}$&1.8146&1.8158&4.1244&1(-3)&1.15&0.004&0.18 \\
\end{tabular} 
\end{table}
\end{minipage}

\begin{figure}[t!]
{\samepage\columnwidth20.5pc
\centering \epsfxsize=8.3cm\epsfbox{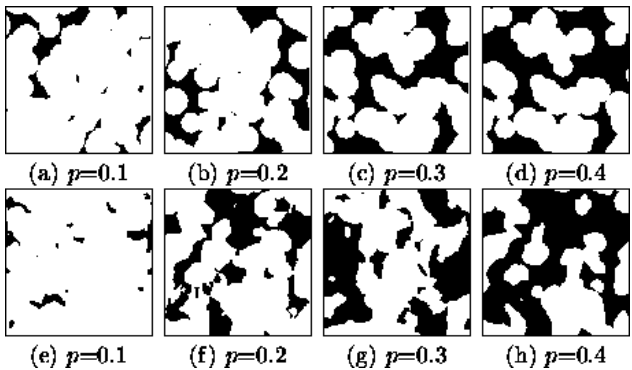}

\vspace{1mm}

\caption{The IOS model (a-d) and reconstructions (e-h) which
reproduce the correlation function (Fig.~\ref{p2IOS}) and
chord-distributions (Fig.~\ref{choIOS}) of the model.
The conducting pore space is shown in black and the images are
10$\times$10$\mu$m.
\label{xsecIOS}}
}
\end{figure}

\begin{figure}[bt!]
{\samepage\columnwidth20.5pc
\centering \epsfxsize=8.0cm\epsfbox{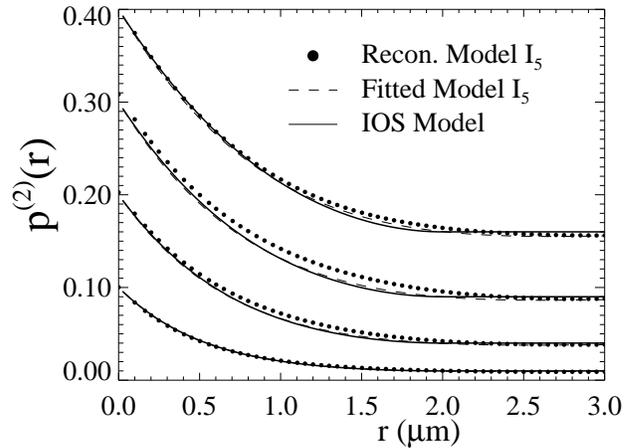}
\caption{The correlation functions of the IOS model compared with
the ``best-fit'' function associated with each reconstruction.
Measurements of $p^{(2)}$ obtained from realisations of the
models are also shown.\label{p2IOS}} 
}
\end{figure}

\begin{figure}[bt!]
{\samepage\columnwidth20.5pc
\centering \epsfxsize=8.0cm\epsfbox{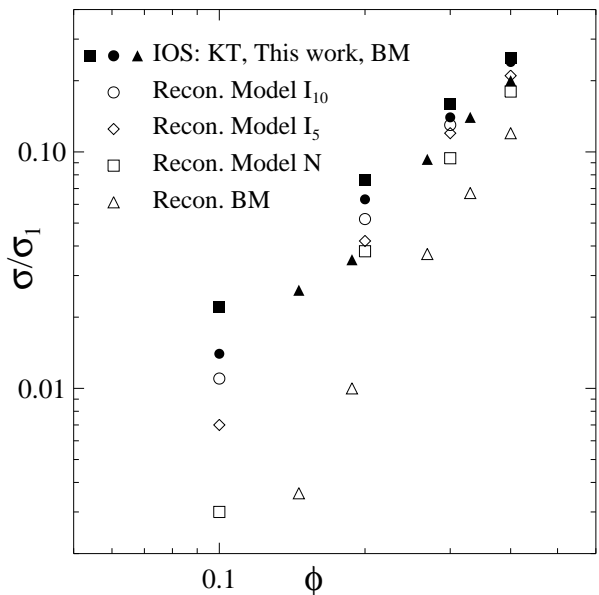}
\caption{Conductivity of the IOS model (solid symbols) compared
with various reconstructions (open symbols). Model I$_{10}$
provides a very good prediction of the actual conductivity.
Other data are from
Refs.~\protect\cite{Kim92} (KT) and~\protect\cite{Bentz94a} (BM).
\label{conIOS}} 
}
\end{figure}

We adopt this strategy to reconstruct the IOS
model at $p\!=\!0.1$, $0.3$ and $0.4$.  In each case models $I_{10}$ and
$I_{5}$ provide the best agreement with the experimental
chord-distributions.  The numerical results are shown in
Table~\ref{tabIOSb} and cross-sections of each model
shown in Fig.~\ref{xsecIOS}.
We have plotted
$p_{fit}^{(2)}(r)$, $p_{exp}^{(2)}(r)$ and measurements of
the function from the reconstructed samples in Fig.~\ref{p2IOS}.
The measured data shows
some deviation from $p_{fit}^{(2)}(r)$ for $p=0.3$. This is due to
the accumulation of errors as we form the intersection sets of
progressively more phase functions. 
Conductivity data is given in
Table~\ref{tabIOSc} and plotted in Fig.~\ref{conIOS}. Models I$_5$
and I$_{10}$ provide a progressively better estimate of the
conductivity. 
We anticipate that increasing the order of model
I$_n$ would yield better estimates. The results indicate that
we have successfully reconstructed the IOS model.

In Fig.~\ref{conIOS} we have also
plotted other data for the IOS model. Kim and
Torquato~\cite{Kim92} (KT)
estimated $\sigma$ for the IOS model using a random
walker algorithm specifically designed to handle locally spherical
boundaries. In the worst case $p=0.1$ our data underestimate that
of KT by a factor of 1.6 (the error decreases
significantly at higher volume fractions). This is probably due to the
discretisation effects of our finite difference
scheme~\cite{Roberts95a}.

\vspace{-4mm}

\noindent
\begin{minipage}{8.5cm}
\begin{table}
\caption{
The conductivity of the IOS model and various reconstructions. 
Models which match the IOS chord-distributions ($I_{5,10}$) provide
better estimates of the conductivity than a reconstruction
based on the single level-cut model (Rec. N).
\label{tabIOSc}}
\begin{tabular}{cccccc}       
$p$ & IOS (KT)\tablenotemark[1]
& IOS & Rec.\ N & Rec.\ I$_5$ & Rec.\ I$_{10}$ \\
\hline
0.1 & 0.022 & 0.014 & 0.003 & 0.007 & 0.011 \\
0.2 & 0.076 & 0.063 & 0.038 & 0.042 & 0.052 \\
0.3 & 0.16  & 0.14  & 0.094 & 0.120 & 0.13  \\
0.4 & 0.25  & 0.24  & 0.180 & 0.210 & 0.22 
\end{tabular} 
\tablenotetext[1]{Kim and Torquato, Ref.~\cite{Kim92}}
\end{table}
\end{minipage}

\vspace{-2mm}

\noindent
This does not alter our
conclusions as all the data presented at a given volume fraction
are presumably effected in the same manner.
The data of Bentz and Martys~\cite{Bentz94a} (BM) for the
IOS model and their one-cut reconstruction are consistently
lower than ours.

\section{Conclusion} 
\label{conclusion}
We have developed a method of reconstructing three-dimensional
two-phase composite materials from information
which can be obtained from digitised micrographs. First a range of
models are generated which share low order
(volume fraction and two-point correlation function) statistical
properties with the experimental sample. The model
which most closely reproduces the chord-distributions of the
experimental material is chosen. The distribution functions
provided a better signature of microstructure than the
three-point correlation function and are simpler to measure than
the microstructure parameters $\zeta$ and $\eta$.
Significantly the three-point and higher order correlation functions 
of the reconstructions can be calculated and employed in
rigourous analytical microstructure-property relationships.

\noindent
Three-dimensional realisations of the models can also be simply generated
for the purpose of numerically evaluating macroscopic properties. 

We found that materials with practically identical two point
correlation functions can have very different morphologies and
macroscopic properties. This demonstrates that
reconstructions based on this information
alone~\cite{Bentz94a,Adler90,Yao93,Joshi74,Quiblier84,Adler92,%
Ioannidis95,Giona96,Losic97}
do not necessarily provide
a useful model of the original material.
If the correlation function exhibits
strong oscillations we found evidence that
prior methods will provide satisfactory reconstructions.
In this case it is important to compare the chord-distributions
of the model and experimental materials.

Our method can be applied to a wider range of composite
and porous media than prior reconstruction techniques.
The generality of the method is achieved by
incorporating new models based on the intersection
and union sets of level-cut GRF models.
The former have recently been
shown to be applicable to organic aerogels~\cite{RobertsAero}
and porous sandstones~\cite{Roberts96c}, while the latter
may be useful for modelling closed-cell foams.
Techniques based on the single-cut GRF
model cannot reproduce the low percolation thresholds
of these materials~\cite{Roberts96a}.
The method was successfully used to reconstruct
several test composites and the overlapping sphere model
over a range of volume fractions. The
reconstructions are better able to model the morphology and
transport properties of the IOS model than prior
studies~\cite{Bentz94a}.

There are several problems with the reconstruction procedure.
First, it is possible that two materials with different properties
may share first and second order statistical information and
chord-distribution functions. In this case the reconstruction method
could fail to yield good estimates of the macroscopic properties.
Second, the generality
of the models we have employed is not sufficient to mimic
all real composites (although prior studies have shown
them to be appropriate for a wide range of
materials~\cite{Roberts96a,Roberts95b%
,Knackstedt95a,Roberts96b,Roberts96c,RobertsAero}).
An example is provided above where our
nine basic reconstructions were unable to model the
chord-distribution of the IOS model. In this case a further
generalisation was found to be successful.
Others are possible. For example, the restriction
that the level-cut and length scales parameters
are identical for each component of the intersection and union sets
can be relaxed, or overlapping spheres can be incorporated in the
level-cut scheme. However the problem remains. It is unlikely,
for example, that the morphology of randomly packed hard spheres
could be mimicked by this scheme.
Third, models formed
from the union and intersection sets contain sharp edges which
are energetically unfavourable in many materials. However there
is little evidence that these play a strong role in determining
macroscopic properties.

New techniques of characterising microstructure are currently
being developed such as those based on
information-entropy~\cite{Andraud97}.
These may contribute to the
problem of selecting the best reconstructions.
Our work also has application to the inverse-problem
of small-angle X-ray scattering from amorphous materials.
In this case the problem is made more difficult
by the absence of higher-order information
such as chord-distributions (although some progress may be
possible~\cite{Levitz92}). Work is underway to model
anisotropic composites and apply the method to
experimental systems.

\acknowledgments
I would like to thank Mark Knackstedt and Dale Bentz for helpful
discussions and the super-computing units at the Australian
National University and Griffith University.

%\bibliographystyle{../Refs/prsty}
%\bibliography{../Refs/grm02,../Refs/dcg02,../Refs/apply02}   

\begin{thebibliography}{10}
\bibitem{prenote2}
The fact that $\rho^{(1)}(r)$ contains high-order statistical information (in
  the sense suggested by the hierarchy of correlation functions) is
  demonstrated by the following argument. It can be shown that
  $\rho^{(1)}(r)=p^{-1}\int_0^\infty r \rho^{(1)}(r) dr \times
  d^2L^{(1)}(r)/dr^2$ where $L^{(1)}(r)$ is the so called lineal-path
  function~\protect\cite{TorqLu93}. This function expresses the probability
  that a line of length $r$ placed randomly in the material will lie entirely
  with in phase one. This quantity is approximately equal to the probability
  that $N$ points along a line lie in phase one, ie.\ it can be obtained from
  the $N$-point function. As the distance between the points shrinks (ie.\
  $N\to\infty$) equality is established.
\bibitem{prenote3}
In part this can be understood in terms of the close relationship between the
  functions (e.g. $\lim_{r\to0} p^{(3)}(r,s,t)\to p^{(2)}(s)$ and
  $\lim_{r,s\to\infty} p^{(3)}(r,s,t)\to p\!\times\! p^{(2)}(t)$). However the
  fact that $\eta$ and $\zeta$ can vary significantly between composites with
  similar $p^{(3)}$ (e.g.\ Table~\ref{tabIIb}) indicates that this function
  contains distinguishing features of the microstructure: these are just not
  seen by the measure $Ep^{(3)}$.
\end{thebibliography}

\end{multicols}

\end{document}